# Convolutional encoding of 60,64,68,72-bit self-dual codes

Alexandre Zhdanov
Voronezh, Russia
a-zhdan@vmail.ru

***Abstract*— In this paper we obtain the [60,30,12], [64,32,12], [68,34,12], [72,36,12] self-dual codes as tailbitting convolutional codes with the smallest constraint length K=9. In this construction one information bit is modulo two added to the one of the encoder outputs and the first row in the quasi-cyclic generator matrix is replaced by the obtained row. The pure quasi-cyclic construction with K=10 is also available for [68,34,12] and [72,36,12] codes. The new [72,36,12] singly even self-dual code with parameters Beta=483 Gamma=0 was obtained.**

## I. Introduction

The convolutional encoding technique for binary self-dual codes was pioneered by G. Solomon and van Tilborg in [1,2,3] The number of famous codes for example (24, 12, 8) extended Golay code, extended (48, 24, 12) and (80, 40, 16) quadratic residue (QR) codes are represented as tailbitting convolutional codes. All of them are quasi-cyclic codes. The quasi-cyclic code is a code for which every cyclic shift of a codeword by $l$ symbols yields another valid codeword. The quasi-cyclic code of $R = 1/m$ consists of $m$ circulants. A circulant is a square matrix where the next row is obtained by one element cyclically shifting the previous row. The tailbitting convolutional code of $R = 1/2$ is a quasi-cyclic code with $l = 2$, where the columns of the circulants mixed to form a compact mixed polynomial string. The mixed polynomial string is a non-zero part of the generator matrix row. This is a trellis code with the same initial and the final encoder states. Those techniques allow implementing Viterbi like decoding algorithm and taking advantage of large free distance with proper decoding complexity. The complexity of Viterbi decoding is $O\left(2^{K-1}\right)$, where $K$ is an encoder constraint length. The aim of the paper is to reduce the constraint length $K$ of the possible tailbitting convolutional encoders for self-dual codes with length 60-72 bits. The known result [4] for extended (72, 36, 12) QR code is $K = 11$.

Self dual codes are a powerful class of codes. Self-dual code $C$ is a code with coding rate $R = 1/2$, where the inner product of any two rows in a generator matrix $G$ gives 0. In other words: $C = C^{\perp}$, where $C^{\perp}$ is a dual code. All codewords of binary self-dual code has even weight. If all codewords weights $\equiv 0 \pmod 4$ the code is called doubly even, if all codewords weights $\equiv 2 \pmod 4$ the code is called singly even. The code is called extremal if the minimum weight of the codeword meets the following bond: $d \le 4\lfloor n/24 \rfloor + 6$ if $n \equiv 22 \pmod{24}$ and $d \le 4\lfloor n/24 \rfloor + 4$ otherwise [5, 6]. We refer the reader to [7] for details. Mainly, the proposed technique [1,2] includes:

1. Pure quasi-cyclic construction (Type $A_0$). In this case the generator matrix is obtained for example by cyclically shift of the mixed polynomial string with step 2 or use another form of generator matrix $G = \left[P \mid Q\right]$, where $P, Q$ are circulants. When one circulant is an identity matrix $I$ the construction $G = \left[I \mid Q\right]$ is called pure double circulant.
2. Adding parity checks to $P$ or/and $Q$ (Type $A_1$).
3. Increasing the code dimension by adjoining to the Type $A_1$ code a new row (Type $A_2$).

The Type $A_2$ and Type $A_1$ constructions are often used simultaneously to obtain a new code. The good example is a bordered double circulant construction:

$$G = \left[\begin{array}{cccc|c} 1 & & & & \\ & 1 & & & \\ & & \ddots & & \\ & & & 1 & \end{array}\right|\left.\begin{array}{ccc} 1 & \cdots & 1\ 0 \\ & & 1 \\ & R' & \vdots \\ & & 1 \end{array}\right],$$

where $R'$ is a circulant of smaller dimension.

Our approach is differs by the following. We use the catastrophic code to obtain the Type $A_0$ code. An "all-ones" input will result a zero weight codeword. Then we replace the first row with a new row which contains the "all-zeros" part for one circulant and the "all-ones" part for other circulant. Then we verify this code by the full weight enumerator analysis. In such a way using the one polynomial set with $K = 9$ and only by changing the length $n$ we can obtain the [60,30,12], [64,32,12], [68,34,12], [72,36,12] self-dual codes. Let us call that coding scheme as Type $A_3$ codes. Also we found the two polynomial sets that provide Type $A_0$ codes with $K = 10$ of length 68. One of them gives the singly even [72,36,12] Type $A_0$ code.

Both give the type $A_2$ [72,36,12] doubly even self-dual codes.

## II. CODE CONSTRUCTION

### A. Type $A_0$ codes

Let us consider the following polynomial pair: $p_1 = [1,0,1,1,1,1,1,0,0,1]$, $p_2 = [1,0,0,1,1,1,1,1,0,1]$. The corresponding tailbitting convolutional encoder structure is shown at Fig. 1.

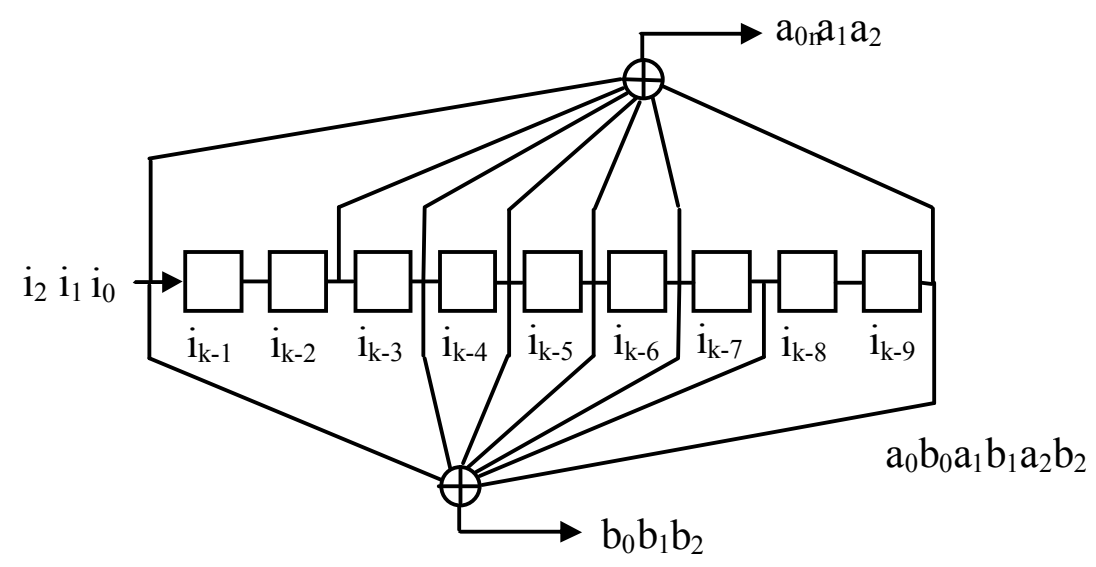

Fig. 1. The encoder structure of type $A_0$ code for p1=$(1371)_8$,p2=$(1175)_8$

The encoder encodes $i_0,\cdots,i_{k-1}$ bits into the sequence $a_0b_0a_1b_1\ldots a_{k-1}b_{k-1}$ of length $n = 2k$. To avoid the tail bits the encoder is initialized by the final state. So iterative decoding over continuously recycling received codeword should be done.

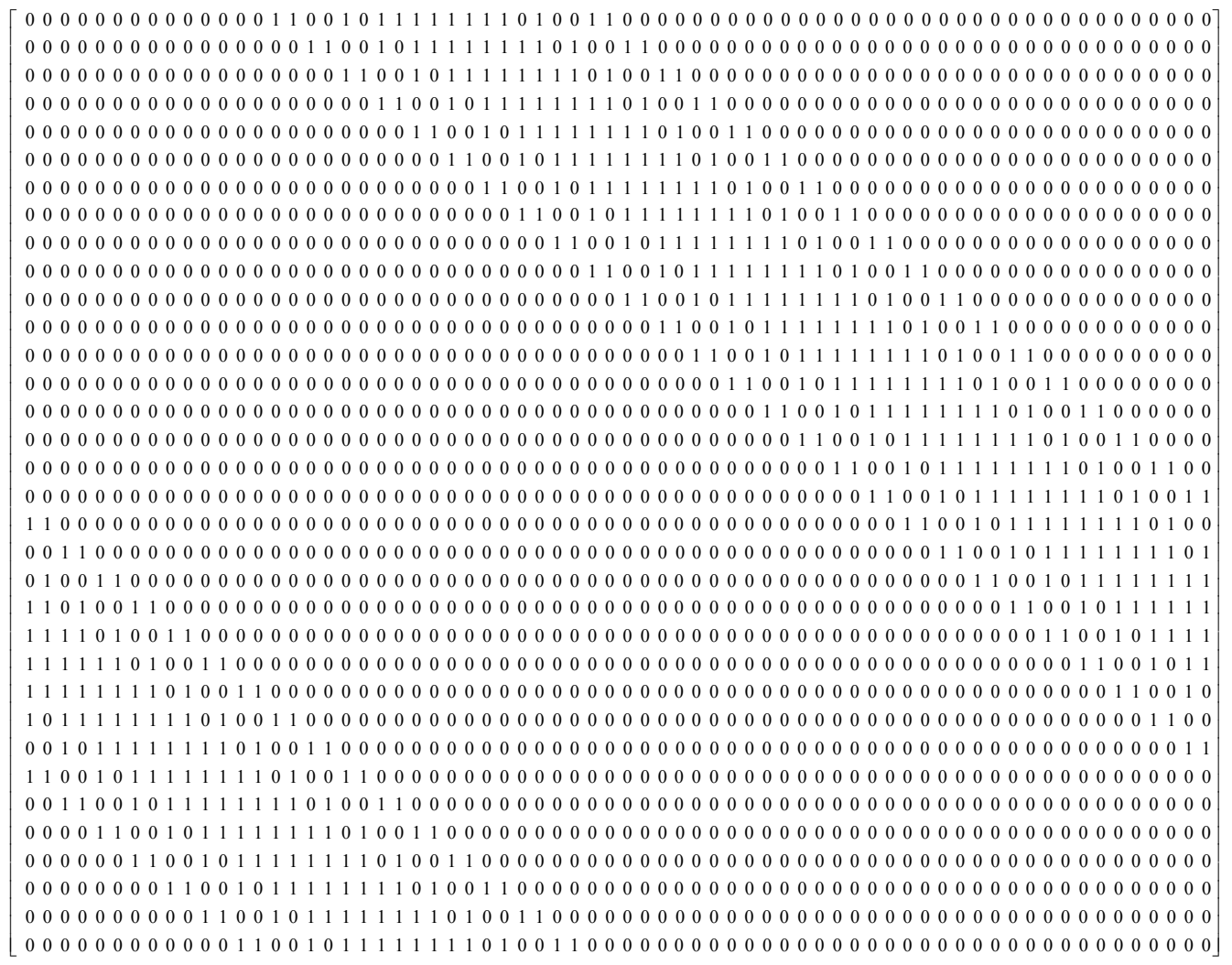

Fig. 2. The generator matrix G of type $A_0$ (n,k) code for n=68, k=34, p1=$(1371)_8$, p2=$(1175)_8$

The corresponding generator matrix G is shown at Fig. 2. The first row of the generator matrix G should be obtained in the following way: the first generator polynomial in binary form is placed on the odd positions of the row, the second generator polynomial in binary form is placed on the even positions of the row, add zeros to complete the row to length $n$. In such a way cyclic shift with step 2 means delay operator for both polynomials.

### B. Type $A_2 - A_1$ codes

The corresponding encoder structure for Type $A_2$-$A_1$ code with polynomials p1=$(1371)_8$, p2=$(1175)_8$ is shown at Fig.3. The encoder encodes $i_1,\cdots,i_{k-1}$ sequence of $k-1$ bits without $i_0$. The b-outputs of the encoder are modulo two added to $i_0$. The $a_{k-1}b_{k-1}$ bits are parity checks.

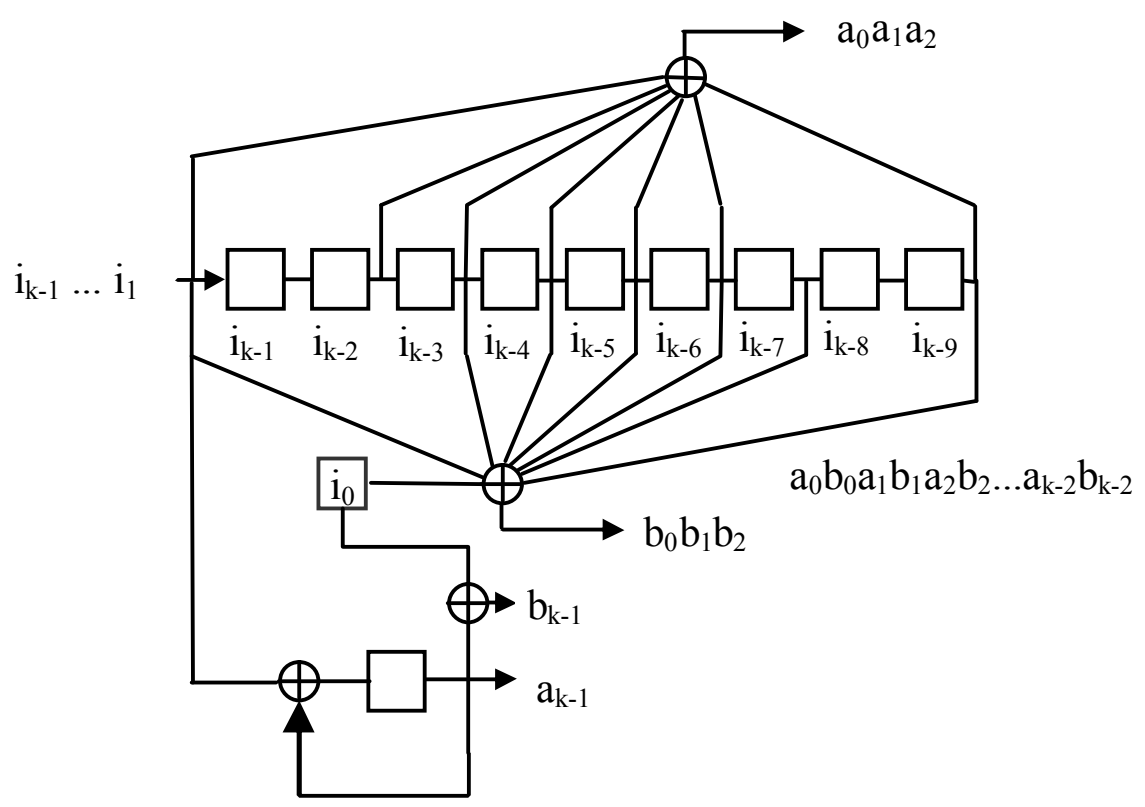

Fig. 3. The encoder structure of type A2-A1 code for p1=(1371)8,p2=(1175)8

The corresponding generator matrix G is shown at Fig. 4.

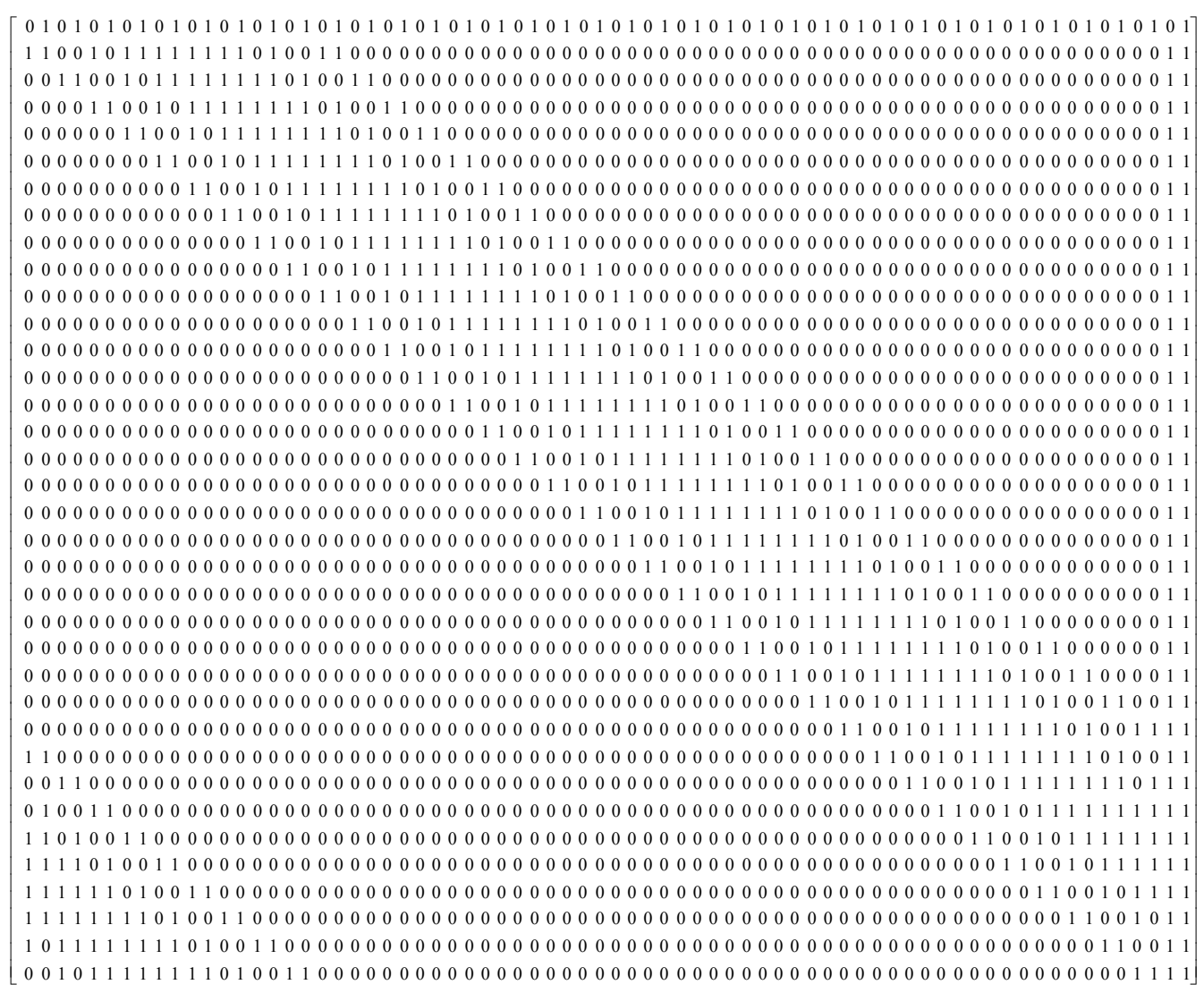

Fig. 4. The generator matrix G of type $A_2$-$A_1$ (n,k) code for n=72, k=36, p1=$(1371)_8$, p2=$(1175)_8$

### C. Type $A_3$ codes (new)

Let us consider the following polynomial pair: $p_1 = [1,1,1,0,0,1,0,1,1]$, $p_2 = [1,1,0,1,0,0,1,1,1]$. The corresponding encoder structure for Type $A_3$ code with polynomials p1=$(713)_8$,p2=$(647)_8$ is shown at Fig.5. The encoder encodes $0,i_1,\cdots,i_{k-1}$ sequence of $k$ bits where the bit $i_0$ is replaced by $0$. The a-outputs of the encoder are modulo two added to $i_0$.The decoding is performed by multiple hypothesis testing. In the first decoding attempt we consider $i_0 = 1$ and in the second decoding attempt we consider $i_0 = 0$. In the $A_2$-$A_1$ case we should test the hypothesis on parity check bits. As we can see pure quasi-cyclic construction with polynomials p1=$(713)_8$, p2=$(647)_8$ is impossible for length 60,64,68,72. An "all-ones" input will result a zero weight output.

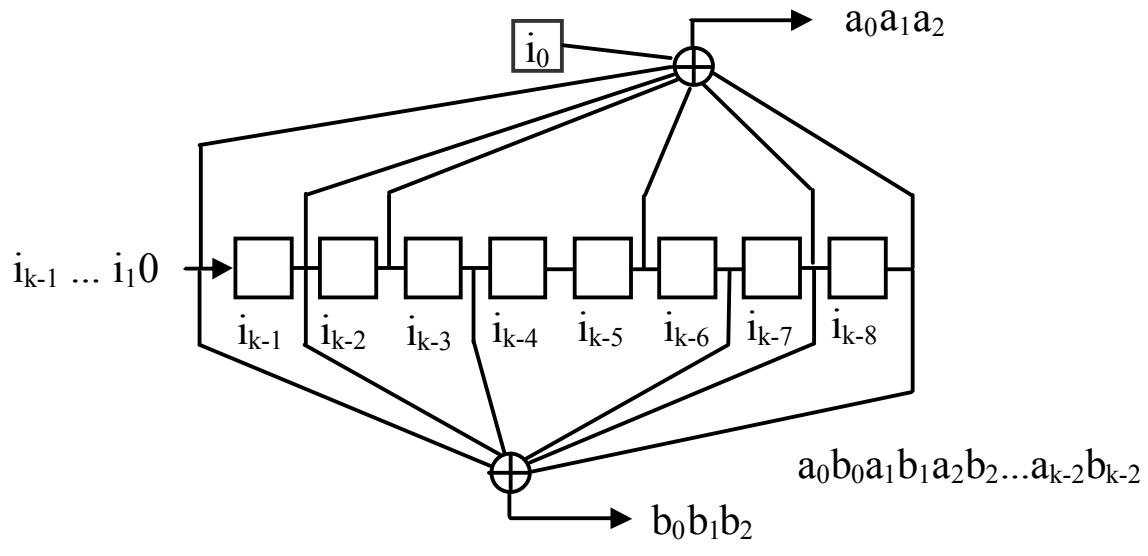

Fig. 5. The encoder structure of type $A_2$-$A_1$ code for p1=$(713)_8$,p2=$(647)_8$

The corresponding generator matrix G is shown at Fig. 6.

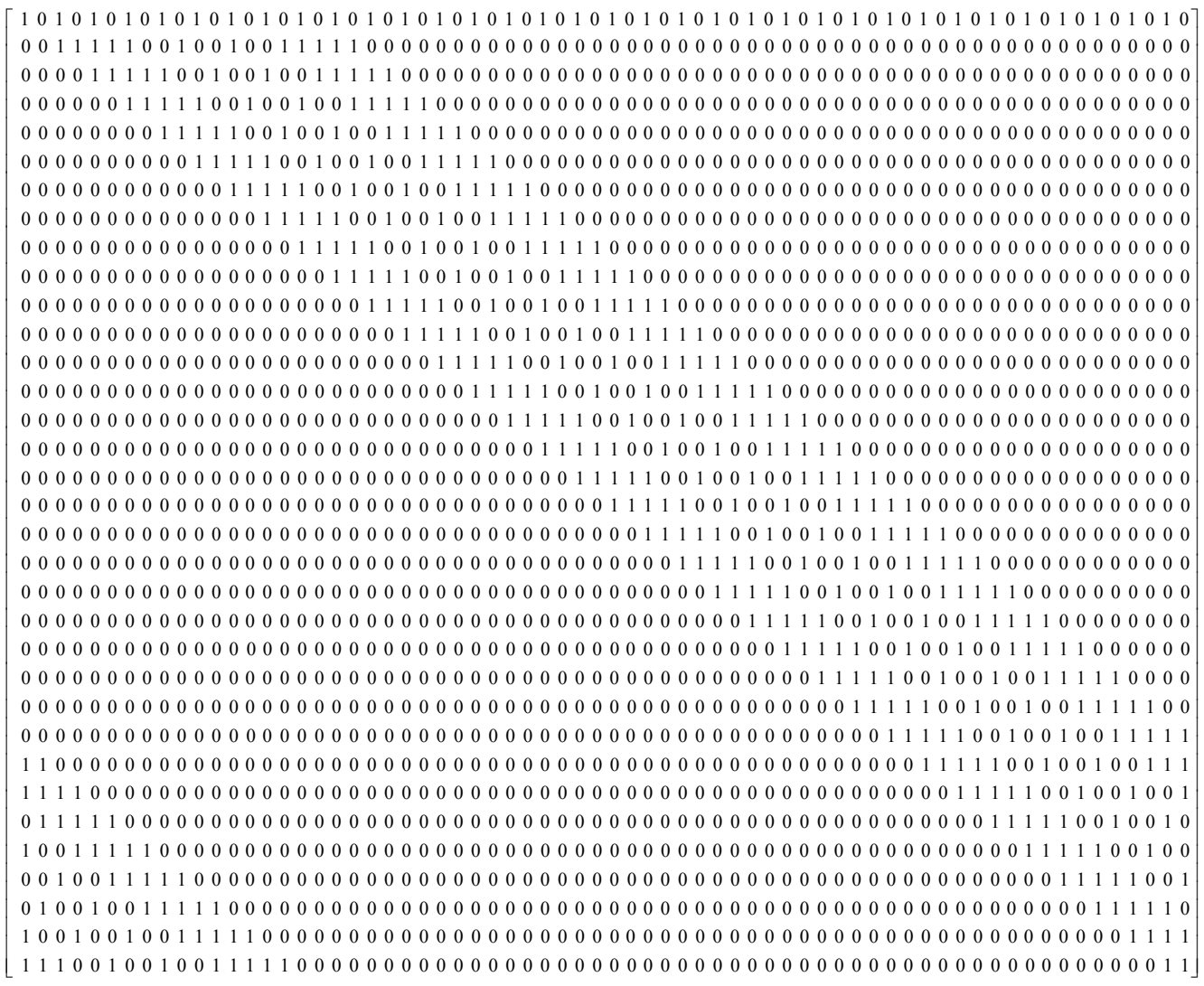

Fig. 6. The generator matrix G of type $A_3$ (n,k) code for n=68, k=34, p1=$(713)_8$, p2=$(647)_8$

## III. Main Result

We have obtained several codes with minimal weight codeword $d = 12$. These codes are listed below.

TABLE I. CODE CONSTRUCTION

| Code | Parameters | | |
|---|---|---|---|
| | *P1* | *P2* | *Type* |
| $C_{n,1}$ | $(1,1,1,0,0,1,0,1,1)_2$ $(713)_8$ | $(1,1,0,1,0,0,1,1,1)_2$ $(647)_8$ | $A_3$ |
| $C_{n,2}$ | $(1,0,0,1,0,1,1,1,1)_2$ $(457)_8$ | $(1,1,1,1,0,1,0,0,1)_2$ $(751)_8$ | $A_3$ |
| $D_{n,1}$ | $(1,0,1,1,1,1,1,0,0,1)_2$ $(1371)_8$ | $(1,0,0,1,1,1,1,1,0,1)_2$ $(1175)_8$ | $A_0$ |
| $D_{n,2}$ | $(1,1,1,1,0,0,1,0,1,1)_2$ $(1713)_8$ | $(1,1,0,1,0,0,1,1,1,1)_2$ $(1517)_8$ | $A_0$ |
| $E_{n,1}$ | $(1,0,1,1,1,1,1,0,0,1)_2$ $(1371)_8$ | $(1,0,0,1,1,1,1,1,0,1)_2$ $(1175)_8$ | $A_2$-$A_1$ |
| $E_{n,2}$ | $(1,1,1,1,0,0,1,0,1,1)_2$ $(1713)_8$ | $(1,1,0,1,0,0,1,1,1,1)_2$ $(1517)_8$ | $A_2$-$A_1$ |

### *A. [60,30,12]*

Only $C_{60,1}$ have weight enumerator $W_{60}$ defined in [8] with parameter $\beta = 0$. The code is singly even code due to the first row.

$$W_{60} = 1 + (2555 + 64\beta)y^{12} + (33600 - 384\beta) + \ldots$$

### *B. [64,32,12]*

Both $C_{64,1}$ and $C_{64,2}$ have weight enumerator defined in [9]. Both codes are doubly even codes.

$$W = 1 + 2976y^{12} + 454956y^{16} + 18\ 275616y^{20} + 233419584y^{24} + \cdots .$$

### *C. [68,34,12]*

The possible weight enumerators for singly even self dual code of length 68 bits defined in [10] are

$$W_{68,1} = 1 + (442 + 4\beta) + y^{12} + (10864\text{-}8\beta)\ y^{14} + \cdots +$$
$$W_{68,2} = 1 + (442 + 4\beta)y^{12} + (14960\text{-}8\beta\text{-}256\gamma)\ y^{14} + \cdots +$$

All obtained codes are listed in the table below. There are singly even codes.

TABLE II. THE 68 BITS LENGTH CODES

| Code | Parameters | | |
|---|---|---|---|
| | $\beta$ | $\gamma$ | *Type* |
| $C_{68,1}$ | 272 | 0 | $W_{68,2}$ |
| $C_{68,2}$ | 238 | 0 | $W_{68,2}$ |
| $D_{68,1}$ | 102 | 0 | $W_{68,2}$ |
| $D_{68,2}$ | 170 | 0 | $W_{68,2}$ |
| $E_{68,1}$ | 170 | - | $W_{68,1}$ |
| $E_{68,2}$ | 137 | - | $W_{68,1}$ |

### *D. [72,36,12]*

The singly even code $D_{72,2}$ has weight enumerator $W_{72,1}$ defined in [11] with parameters $\beta = 483$ and $\gamma = 0$.

$$W_{72,1} = 1 + 2\beta y^{12} + \left(8640 - 64\lambda\right)y^{14} + \left(124281 - 24\beta + 384\gamma\right)y^{16} + \cdots$$
$$W_{72,2} = 1 + 2\beta y^{12} + \left(7616 - 64\gamma\right)y^{14} + \left(134521 - 24\beta + 384\gamma\right)y^{16} + \cdots$$

The code with these parameters is not known to exist.

The possible weight enumerator for doubly even self dual code of length 72 bits defined in [12] is

$$W = 1 + (4398+\alpha)y^{12} + (197073-12\alpha)y^{16} + (18396972+66\alpha)y^{20} + \cdots$$

The code $C_{72,1}$ has $\alpha = -2820$. The code $C_{72,2}$ has $\alpha = -2748$. Both $E_{72,1}$ and $E_{72,2}$ have $\alpha = -3096$.

## IV. Conclusion

The obtained codes with the smallest constraint length K=9 are good tradeoff of performances-complexity for practical encoder and decoder implementation. The code $D_{72,2}$ has not been listed in [11,13,14,15] and it is the first known code with weight enumerator $W_{72,1}$ and $\beta = 483 \quad \gamma = 0$.